\documentclass[conference]{IEEEtran}
\IEEEoverridecommandlockouts
\usepackage{url}\usepackage{subfig}
\usepackage[ruled,linesnumbered]{algorithm2e}
\SetKwInput{KwInput}{Input}                
\SetKwInput{KwOutput}{Output}              
\newcommand{\nonl}
{\renewcommand{\nl}{\let\nl\oldnl}}
\usepackage{enumitem}
\usepackage{multirow,array}
\usepackage{cite}
\usepackage{graphicx}
\usepackage{psfrag}
\usepackage{array}
\usepackage{algorithm2e}
\usepackage{amssymb}
\usepackage{amsfonts}
\usepackage{float}
\usepackage{textcomp}
\usepackage{xcolor}
\usepackage{tabu}
\usepackage{array}
\usepackage{tabularx}
\usepackage{subfiles}
\usepackage{comment}
\usepackage{amssymb}
\usepackage{pifont}
\usepackage[amsthm,thmmarks]{ntheorem}

\newcolumntype{P}[1]{>{\centering\arraybackslash}p{#1}}
\newcolumntype{M}[1]{>{\centering\arraybackslash}m{#1}}

 \usepackage{amsmath}
 \usepackage{blkarray, bigstrut}
\usepackage{mathtools}
\ExplSyntaxOn
\NewDocumentCommand{\RN}{m}
 {
  \textup{ \int_to_Roman:n { #1 } }
 }
\ExplSyntaxOff

\usepackage{nicematrix}

\def\BibTeX{{\rm B\kern-.05em{\sc i\kern-.025em b}\kern-.08em
    T\kern-.1667em\lower.7ex\hbox{E}\kern-.125emX}}
\usepackage{float}

\newtheorem{conjecture*}{Conjecture}
\newtheorem{lemma}{Lemma}

\newtheorem{corollary}{Corollary}
\newtheorem{remark}{Remark}
\newtheorem{definition}{Definition}
\newtheorem{theorem}{Theorem}
\newtheorem{example}{Example}

\newcommand{\bts}{\color{blue}}

\newcommand{\mc}{\mathcal}
\newcommand{\D}{\Delta}

\newcommand{\tth}{\texttt{th}}
\newcommand{\ch}{\mathcal{H}}

\title{A Hypergraph based lower bound on Pliable Index Coding based on Nested Side-Information Sets%
\thanks{This paper has been published as part of the 2025 XIX International Symposium on Problems of Redundancy in Information and Control Systems (Redundancy), IEEE.}}

\author{
  \IEEEauthorblockN{Tulasi Sowjanya B., Prasad Krishnan}
  {
                    }
\vspace{-0.2cm}
}
\begin{document}
\maketitle
\setlist[itemize]{leftmargin=*}


\begin{abstract}
In pliable index coding (PICOD), a number of clients are connected via a noise-free broadcast channel to a server which has a list of messages. Each client has a unique subset of messages at the server as side-information, and requests for any one message not in the side-information. A PICOD scheme of length $\ell$ is a set of $\ell$ encoded transmissions broadcast from the server such that all clients are satisfied. Finding the optimal (minimum) length of PICOD and designing PICOD schemes that have small length are the fundamental questions in PICOD. In this paper, 
we present a new lower bound for the optimal PICOD length using a new structural parameter called the nesting number, denoted by  $\eta(\ch)$ associated with the hypergraph $\ch$ that represents the PICOD problem. While the nesting number bound is not stronger than previously known bounds, it can provide some computational advantages over them. Also, using the nesting number bound, we obtain novel lower bounds for some PICOD problems with special structures, which are tight in some cases. 
\end{abstract}
\let\thefootnote\relax\footnotetext{
 Tulasi and Dr.\ Krishnan are with the Signal Processing \& Communications Research Center, International Institute of Information Technology, Hyderabad, India. Emails:\{tulasi.sowjanya@research., prasad.krishnan@\}iiit.ac.in. A longer version of this work , with additional sections containing other results, is under review in the IEEE Transactions on Information Theory.}

\section{Introduction}
The problem of \textit{pliable index coding (PICOD)}, introduced in \cite{brahma2015pliable} by Brahma and Fragouli,  consists of a server with messages, and a number of clients, connected via a noiseless broadcast channel. Each client possesses a subset of messages as \textit{side-information} and demands for \textit{any} message from the set of messages in its \textit{request set} (those messages not present as its side-information). A PICOD problem can thus be defined by the collection of unique side-information sets or request sets of all the clients. The server then designs (possibly coded) transmissions and broadcasts them to the clients, which then decode their desired symbols. This is called as a pliable index coding scheme, or a \textit{PICOD scheme}. The number of transmissions made by the server is termed as the \textit{length} of the PICOD scheme, and the goal is to design PICOD schemes with lengths as small as possible. The PICOD problem is a variant of the well-studied index coding problem, introduced in \cite{ISCOD_BiK} by Birk and Kol. The setting in index coding is different from PICOD only in the sense that each client demands a specific message in the request set, rather than any message. Newer variants of PICOD have also appeared in recent literature, such as the preferential PICOD (PPICOD) \cite{byrne2023preferential}. In this, the clients have varying preferences among the unknown messages, and wish to receive one with a higher preference. Apart from being a canonical problem in information theory, ideas from pliable index coding prove useful for constructing data-shuffling schemes for distributed computing \cite{song2019pliable_shuffling} and private information retrieval \cite{obead_PPIRITW}.

In \cite{brahma2015pliable}, it was shown that finding the optimal code length for PICOD problem is NP-hard.
They have also presented few algorithms in \cite{brahma2015pliable}.
In \cite{8036234}, Song and Fragouli presented a deterministic algorithm for PICOD termed  \textsf{BinGreedy}, which constructed a PICOD scheme with at most $\mc{O}(\log^2n)$ transmissions. An improved greedy cover algorithm titled \textsf{ImpGrCov}, was presented in \cite{eghbal2023improved}. 
In \cite{10403947}, Krishnan et al. presented a hypergraph coloring approach to the PICOD problem. 
Using several variants of \textit{conflict-free colorings} \cite{even2003conflict} of the PICOD hypergraph, randomized algorithms were presented for PICOD and PICOD($t$) problems.
A simple greedy algorithm for PICOD, which uses the hypergraph representation of the PICOD problem was presented in \cite{subramanian2022bounding}. This gives an achievable linear scheme with length at most $\D(\mathcal{H})$, where $\D(\mathcal{H})$ is the maximum degree of any vertex in the hypergraph $\ch$. 

Information theoretic converses for various special classes of PICOD problems were derived in \cite{liu2019tight,8849812_ShanuRaj_CodeConstrPIC}. Specifically, the work \cite{liu2019tight} considered a class of problems known as complete-$\Sigma$ PICOD($t$), where $\Sigma\subseteq[0:m-1]$ defines the side-information sets of the clients present in the problem. Tight converses for complete-$\Sigma$ PICOD($t$) for a number of special choices for $\Sigma$ were presented in \cite{liu2019tight}, utilizing the notion of \textit{decoding chains} and earlier converse results for index coding \cite{neely2013dynamic}. The work \cite{ong2019optimal} generalized the techniques in \cite{liu2019tight}, and obtained lower bounds on the optimal PICOD length for general PICOD problems. This was later improved in \cite{ong2019improved}. The bounds in \cite{ong2019optimal,ong2019improved} are also based on the idea of decoding chains. They remain the only known bounds for general PICOD problems; however, they could be difficult to compute in general.

In the present work, we consider a slightly different hypergraph model for PICOD that was used in \cite{10403947} and \cite{subramanian2022bounding}. The vertex set remain the same in both the models, but the edges represent side-information sets of clients rather than representing the request sets. New converse results for PICOD based on this model are presented here.

The paper is organized as follows. We  present the formal PICOD system model and the hypergraph framework in Section \ref{sec:systemmodel_and_hypergraph}. We then briefly review the existing converses for general PICOD problems from \cite{ong2019optimal,ong2019improved} (Subsection \ref{subsec:conversesabsentreceiver}). In Subsection \ref{subsec:nesting}, we identify a new structure called a \textit{nested collection} of the side-information sets of clients, and an associated parameter we call the \textit{nesting number} $\eta(\ch)$. Via the bound from \cite{ong2019optimal}, we show that the optimal PICOD length is lower bounded by $\eta(\ch)$ (Theorem \ref{thm:lowerboundeta}). In Subsection \ref{subsec:algo_nesting number-revised}, we provide a polynomial-time algorithm (Algorithm \ref{algo:nestedcollectionforspecificROOTclient}) to construct nested collections of clients, each of which provides a lower bound for $\eta(\ch)$. In Subsection \ref{subsec:applicationsofnestingnumber}, we show a class of PICOD problems for which $\eta(\ch)$ is a tight lower bound (Lemma \ref{lemma:specificclassEtaTight}). Using Theorem \ref{thm:lowerboundeta}, we also prove the converse bound (Lemma \ref{lemma:lowerboundCompleteSproblems}) for complete-$\Sigma$ PICOD problems with general $\Sigma$, re-proving the result from prior work \cite{liu2017ITW}. While the $\eta(\ch)$ lower bound is not stronger than the bound in \cite{ong2019optimal}, nevertheless we show in Subsection \ref{subsec:applicationsofnestingnumber} that it recovers results from \cite{ong2019optimal} involving PICOD problems with various structural constraints (Lemmas \ref{lemma:maxlengthnestedabsent}, \ref{lemma:unionofabsentRxsNOT0-m-1}, \ref{lemma:1or0absentnestedpair}).  Using our results, we also obtain an exhaustive characterization of the optimal PICOD length of all problems upto three clients, along with the associated lower bounds in each case (Lemmas \ref{lemma:twoclientcase} and \ref{lemma:threeclients}). In the process, we see that the converse bounds from Theorem \ref{thm:lowerboundeta} and from \cite{ong2019optimal} can be equally loose.  The paper concludes in Section \ref{sec:discussion} with some directions for further research. Due to page limitations, all the proofs are omitted here and are included in the appendices.

\textit{{Some notation and basic definitions:}} We set up some notation and review some basic definitions related to (hyper)graphs that are useful for this work. For some positive integers $m$ and $n$, $n > m$ we denote the set $\{1, 2, \hdots ,n\}$ by $[n]$ and the sequence $(m, m+1, m+2, \hdots,n)$ by $[m : n]$. For sets $A$ and $B$,  $A \backslash B$ denotes the set of elements in $A$ but not in $B$. If $B=\{v\}$, then we also denote $A\setminus B$ as $A\setminus v$. The notation $\lvert A \rvert$ represents the size of set $A$ and  $\emptyset$ denotes an empty set. $\mathbb{F}_q$ denotes the finite field with $q$ elements. 
We now present some definitions related to hypergraphs. A hypergraph $\mc{H}(\mc{V},\mc{E})$ consists of a pair of sets: the \textit{vertex set} denoted by $\mc{V}$, and the \textit{edge set} denoted by $\mc{E}$. The elements of $\mc{E}$ are subsets of the vertex set. We shall also use $\mc{V}(\mc{H})$ and $\mc{E}(\mc{H})$ to refer to the set of vertices $\mc{V}$ and set of hyperedges $\mc{E}$ of a hypergraph $\mc{H}(\mc{V},\mc{E})$, respectively. 
Two vertices $v_i$ and $v_j$ of $\mc{H}$ are said to be \textit{adjacent} to each other if there is at least one hyperedge which contains both the vertices. 

\section{System Model for PICOD and hypergraph representation}
\label{sec:systemmodel_and_hypergraph}
The system model of pliable index coding (PICOD) problem consists of a server with \textit{m} messages from a finite field, denoted as $b_i:i\in [m]$, and \textit{n} clients connected to the server by a noise-free broadcast link. The set of indices of the side-information messages at client $i$ is given by $S_i \subseteq [m]$. The collection of \textit{side-information} sets is denoted as ${\cal S}=\{S_i:i\in[1:n]\}$. \textit{Request set} of client $i$ is then denoted as$R_i = [m]\mathbin{\setminus} S_i$.


The client $i$ then requires one message from the set of messages not in their side-information sets $\{b_j : j \notin S_i\}$. Note that we assume $\lvert S_i\rvert \neq m, \forall i$. A pliable index code (or a PICOD scheme) for this PICOD problem consists of (a) an encoding scheme at the server which encodes the messages $b_j:j\in[m]$ into $\ell$ symbols from $\mathbb{F}_q$ which the server then broadcasts, and (b) decoding functions at the clients which enable each of them to decode one message from its request set, using the encoded symbols received and its side-information. Here, the quantity $\ell$ is called as the PICOD \textit{length}. For some encoding scheme, a client is said to be \textit{satisfied} if it is able to decode one message from its request set, else it is said to be \textit{unsatisfied}. The goal of pliable index coding is to design PICOD schemes that have length $\ell$ as small as possible, while satisfying all clients. Observe that we can identify client $i$ with its side-information set $S_i$, as any two clients with the same side-information set are essentially identical from the PICOD perspective.

We now briefly describe the hypergraph model for PICOD. A given PICOD problem can be completely represented by a hypergraph $\mc{H}$ with the vertex set $\mc{V} = [m]$ and edge set $\mc{E} = \{S_i \subseteq [m]: i \in [n]\}$. Naturally, every hypergraph also defines a corresponding PICOD problem. We, thus, do not distinguish between hypergraphs and PICOD problems in this work. 
We denote by $\beta_q(\mc{H})$ the smallest length of any PICOD scheme for $\mc{H}$ over $\mathbb{F}_q$. Also, let $\beta(\mc{H}) \triangleq \min_{q}\beta_q(\mc{H})$ denote the optimal length of any PICOD scheme for $\mc{H}$ over any finite field. 

\section{A new lower bound on $\beta(\ch)$ based on nested side-information sets}
\label{sec:lowerbound_via_nesting}
In this section, we derive a new lower bound on optimal length $\beta(\ch)$ based on a new parameter of the hypergraph called the \textit{nesting number}, $\eta(\ch)$. First, we recall the `absent clients'-based converse bounds for the minimum PICOD length of general PICOD problems, available in \cite{ong2019optimal} (and later improved in \cite{ong2019improved}). Subsequently, we prove a new lower bound on the optimal length of PICOD based on subsets of side-information sets with some special structure. We obtain its relationship to the existing converses, and provide new proofs for some existing results based on our lower bound. We also show a class of problems for which our lower bound is tight and derive new results for a general class of complete-$\Sigma$ PICOD problems \cite{liu2019tight}. We further characterize the optimal length of PICOD for upto three clients. 

\subsection{The converses from \cite{ong2019optimal,ong2019improved}}
\label{subsec:conversesabsentreceiver}
We now briefly recount essential ideas regarding the converse bounds for general PICOD problems from prior works \cite{ong2019optimal,ong2019improved}. These converse bounds were essentially based on the earlier converse for index coding from \cite{neely2013dynamic}. 

Let $D$ denote a decoding choice function for the PICOD problem specified by $\mc{H}$, where $D(S_i)\in R_i:i\in[n]$ denotes the decoding choice of client $i$. The converse bounds in both the works \cite{ong2019optimal,ong2019improved} involve constructing a total ordering of the $m$ message indices, for each decoding choice function $D$. Let this totally ordered set be denoted by $C_D$, where the set of first $k$ elements of $C_D$ is denoted by $C_{D,k}$, where $C_{D,k}=C_D$. We recall some terminology from \cite{ong2019optimal}. A subset $S\subsetneq [1:m]$ is said to be an \textit{absent client} of $\ch$, if it is not the side-information set of any client in $\ch$. In contrast, a \textit{present} client refers to a side-information set of a client in $\ch$. In the work \cite{ong2019optimal}, the ordered set $C_D$ is constructed one message at a time, according to the following rules.
\begin{itemize}[leftmargin=0.05in]
    \item \textit{Rule A:} If $C_{D,k}$ is a present client, then the $(k+1)^{\tth}$ message index added to the ordered set is $D(C_{D,k})$, i.e., $C_{D,k+1}=\{D(C_{D,k})\}\cup C_{D,k}$. Such a client $C_{D,k}$ is said to be `hit' by $C_D$.
    \item \textit{Rule B:} Otherwise, if $C_{D,k}$ is an \textit{absent client}, the $(k+1)^{\tth}$ message index added to the ordered set is any message index not already in $C_{D,k}$. Such message indices are said to be `skipped' by $C_D$. 
\end{itemize}
Let $\tau(C_D)$ represent the number of messages added to $C_D$ according to \textit{Rule A}, i.e., $\tau(C_D)$  is the number of clients hit by $C_D$ or the number of \textit{non-skipped} message-indices in $C_D$. Then, the converse bound in \cite[Lemma 5]{ong2019optimal} says that $\beta(\mc{H})\geq \min\limits_D \max\limits_{C_D}\tau(C_D),$ where the minimization is over all possible decoding choice functions for the given problem, and the maximization is over all possible ordered sets $C_D$ chosen for a given $D$, according to the rules above.

In \cite{ong2019improved}, the above rules for constructing $C_D$ are refined further to obtain a improved converse bound. While \textit{Rule A} is kept as is, \textit{Rule B} is replaced with two options. That is, if $C_{D,k}$ is an absent client, then we execute one of the following options as per \cite[Algorithm 1]{ong2019improved}.
\begin{itemize} [leftmargin=0.05in]
    \item \textit{Option B':} If there exists some present client $S\in{\cal S}$ such that $D(S)\notin C_{D,k}$ and $S\subsetneq C_{D,k}$, then we may add $D(S)$ to the ordered set, i.e., $C_{D,k+1}=\{D(S)\}\cup C_{D,k}$. 
    \item \textit{Option C':} Alternatively, let the $(k+1)^{\tth}$ message index added to the ordered set be any message index not already in $C_{D,k}$.
\end{itemize}
Let $\tau'(C_D)$ denote the number of messages added according to \textit{Rule A} or \textit{Option B'}, i.e. the number of non-skipped message-indices of $C_D$. The converse bound in \cite[Lemma 1]{ong2019improved} is $\beta(\mc{H})\geq \min\limits_D \max\limits_{C_D}\tau'(C_D),$ which is in general an improvement over the bound from \cite{ong2019optimal}. This is clear, as we observe that the action in \textit{Option C'} is identical to \textit{Rule B}. Thus, executing only \textit{Option C'} (and never\textit{ Option B'}) whenever $C_{D,k}$ is an absent client, results in the rules for forming $C_D$ as per \cite{ong2019optimal}. Thus, we have
\begin{align}
\label{eqn:oldconversebounds}
    \beta(\mc{H})\geq \tau_2(\ch)\triangleq \min\limits_D \max\limits_{C_D}\tau'(C_D)\geq \tau_1(\ch)\triangleq\min\limits_D \max\limits_{C_D}\tau(C_D).
\end{align}
From the results of \cite{ong2019improved}, the inequality $\tau_2(\ch)\geq \tau_1(\ch)$ can be strict.

\subsection{A lower bound for PICOD via nested collections of $\ch$} 
\label{subsec:nesting}
We now give the definition of a nested collection of clients of the PICOD problem $\ch$, based on its collection $\cal S$ of side-information sets. 

\begin{definition}
\label{defn:nestingcollection}
Consider a PICOD problem given by $\mc{H}$. An ordered list of $L$ subsets of the side-information sets of $\ch$ given by ${\cal S}_i\subseteq \mc{S}: i\in[L]$ is said to be a \textit{nested collection} of $\ch$ if it satisfies the following property: 
    \begin{itemize}[leftmargin=0.05in]
    \item For each $i\in[L-1]$, each $S\in \mc{S}_i$, and each $j\notin S$, there exists some $S'\in \mc{S}_{i+1}$ such that $S \cup \{j\}\subseteq S'$.
\end{itemize}
 We refer to the set $\mc{S}_i$ as the \textit{level-$i$} side-information sets (or equivalently, level-$i$ clients) of the nested collection. We call the number of levels $L$ in such a collection as its \textit{nesting length}. We define the maximum nesting length of any nested collection in $\mc{H}$ as the \textit{nesting number of $\ch$}, and denote it by $\eta({\mc{H}})$.
\end{definition}
\begin{remark}
\label{remark:nestedcollectionviarequestsets}
In a previous version of this work \cite{subramanian2022bounding}, we defined the quantity $\eta(\ch)$ differently, based on the request sets of $\ch$. However, on reflection, we found the definition in \cite{subramanian2022bounding} too restrictive. Our attempts to generalize the earlier idea resulted in Definition \ref{defn:nestingcollection}. 
\end{remark}

Our main result here is the following lower bound for the optimal PICOD length (proof available in the Appendix \ref{subsec:proofoftheoremlowerbound}). 
\begin{theorem}
\label{thm:lowerboundeta}
  Let $\mc{H}$ be a PICOD hypergraph. Then, $\beta(\mc{H}) \geq \tau_2(\ch)\geq \tau_1(\ch)\geq \eta(\mc{H})$.
\end{theorem}



The following Example \ref{example2} illustrates the quantity $\eta(\ch)$ as well as the lower bound in Theorem \ref{thm:lowerboundeta}.
\begin{example}
\label{example2} 

Consider a PICOD problem $\ch$ with four messages indexed by $\{1, 2, 3, 4\}$, and $11$ clients $\mc{S}=\{S_1, \hdots, S_{11}\}$ with their side-information sets being  $S_1 = \emptyset$, $S_2 = \{1, 2\}$, $S_3 = \{3\}$, $S_4 = \{1, 4\}$, $S_5=\{1, 2, 3\}$, $S_6=\{1, 2, 4\}$, $S_7=\{1, 3, 4\}$, $S_8=\{2\}$, $S_9=\{2, 4\}$, $S_{10}=\{2,3\}$ and $S_{11}=\{1, 3\}$. Consider the ordered list of subsets of side-information sets given by $\mc{S}_1=\{S_1\}$, $\mc{S}_2=\{S_2, S_3, S_4\}$, and $\mc{S}_3=\{S_5, S_6, S_7\}$. It is not difficult to verify this is a nested collection. For instance, consider the client $S_2=\{1,2\} \in \mc{S}_2$. For this client, for every $j \notin S_2$, we see that there exists a client in $S^\prime \in \mc{S}_3$ such that $S^\prime \subseteq S \cup {j}$. 

We also observe that there cannot be a nested collection of size $4$ for this problem (this actually requires all clients with side-information sets of sizes $\{0,1,2,3\}$ to be present in $\ch$). Thus, by the definition of $\eta(\ch),$ we have the nesting number $\eta(\ch)=3$.

We now demonstrate a linear PICOD scheme with three transmissions. Consider the collection of transmissions $b_1$, $b_2+b_4$ and $b_1+b_2+b_3$. The first transmission satisfies the clients $S_1$, $S_3$, $S_8$, $S_9$ and $S_{10}$, as message $b_1$ lies in their request sets. The second transmission satisfies the clients $S_2$, $S_4$, $S_5$, and $S_7$, as each of them have precisely one of $b_2$ or $b_4$ in their side-information set. Finally, the last transmission satisfies the remaining clients $S_6$ and $S_{11}$. Thus $\eta(\ch)=\beta(\ch)=3$, for this problem. \hfill $\blacksquare$

\end{example}
\subsection{An algorithm for finding nested collections} 
\label{subsec:algo_nesting number-revised}
Though the $\eta(\ch)$ bound is not stronger than the $\tau_1(\ch)$ bound, in terms of computability, the nested-collections-based bound $\eta(\ch)$ may have advantages in some scenarios. Algorithm \ref{algo:nestedcollectionforspecificROOTclient} gives a method to construct a nested collection rooted at a particular client (i.e., at the level-$1$ of the collection). By running Algorithm \ref{algo:nestedcollectionforspecificROOTclient} with each client as the root, we get several nested collections, some of which can potentially have lengths close to the nesting number. We show that this can be done in time that is polynomial in the system parameters. However, the time-complexity of the algorithms in \cite{ong2019optimal,ong2019improved} which enable computation of the lower bounds in \eqref{eqn:oldconversebounds} are not explicitly specified. These algorithms likely have a larger complexity than our computations of nested collections via Algorithm \ref{algo:nestedcollectionforspecificROOTclient}, as they entail running through all possible choices for the demands $D$, and all valid totally ordered sets $C_D$ as per the rules given in Subsection \ref{subsec:conversesabsentreceiver}.

\begin{algorithm}[htbp]

\small
    \DontPrintSemicolon
     \KwInput {PICOD problem $\mc{H}$ with client side-information sets $\mc{S}=\{S_1,\hdots,S_n\}$, and a specific client-index $a\in[1:n]$. } 
     \KwOutput{A nested collection rooted at $S_a$ and its nesting length $\lambda$.}
     \SetKwRepeat{Do}{do}{while}
    \textbf{Initialize:} ${\cal S}_1\gets \{S_a\}$, $\lambda\gets 1$, $\mathsf{MaxDepthReached}\gets 0$, $i\gets 1$.\;
    \Do{$\mathsf{MaxDepthReached}= 0$}{ 
    $\mc{S}_{\mathsf{temp}}\gets \emptyset$, $\mc{S}_{i+1}\gets \emptyset$. \;
    \For{$S \in \mc{S}_i$}{
    \For{$j \notin S$}{
    $\mc{S}_{j,S}\gets \{S''\in \mc{S}~\colon S\cup \{j\}\subset S''\}.$\;
    \eIf {$\mc{S}_{j,S}\neq \emptyset$}
    {
    $S'\gets \arg\min\limits_{S''\in \mc{S}_{j,S}} |S''|.$\;
    $\mc{S}_{\mathsf{temp}}\gets\mc{S}_{\mathsf{temp}}\cup\{S'\}.$ \;
    }
    {$\mathsf{MaxDepthReached}\gets 1$, $\lambda\gets i$.\;
    \textbf{break}\;}
    }
    \If {$\mathsf{MaxDepthReached}= 1$}
    {\textbf{break}\;}
    }
    \If {$\mathsf{MaxDepthReached}= 0$}
    {
     $\mc{S}_{i+1}\gets \mc{S}_{\mathsf{temp}}$\;
     $i\gets i+1\;$
    }
    }
       Return $\mc{S}_i:i\in[1:\lambda]$ and $\lambda$.
    \caption{Computing a nested collection rooted in a given client (side-information set)} 
    \label{algo:nestedcollectionforspecificROOTclient}

    \end{algorithm}

Algorithm \ref{algo:nestedcollectionforspecificROOTclient} is mostly self-explanatory; we add only a few statements for exposition. Line 6 of the algorithm defines a search-space for the `successor' side-information sets of the presently considered side-information set $S$ (in level-$i$ of the nested collection being formed), and a candidate requested message $j$.  If this set is non-empty, then one among the smallest successors will be added to the next level (lines 8-9), else a flag variable $\mathsf{MaxDepthReached}$ is set to $1$ (line 11), indicating that the maximal length (denoted by $\lambda$) has been reached in this case. This results in the exit out of the loops in the algorithm, at which point the nested collection and its length $\lambda$ is returned. The following example illustrates the working of Algorithm \ref{algo:nestedcollectionforspecificROOTclient}.

\begin{example}
\label{example3} 
Consider a PICOD problem $\ch$ in Example \ref{example2}, with four messages indexed by $\{1, 2, 3, 4\}$, and $11$ clients $\mc{S}=\{S_1, \hdots, S_{11}\}$ with their side-information sets being  $S_1 = \emptyset$, $S_2 = \{1, 2\}$, $S_3 = \{3\}$, $S_4 = \{1, 4\}$, $S_5=\{1, 2, 3\}$, $S_6=\{1, 2, 4\}$, $S_7=\{1, 3, 4\}$, $S_8=\{2\}$, $S_9=\{2, 4\}$, $S_{10}=\{2,3\}$ and $S_{11}=\{1, 3\}$. We now run through the Algorithm \ref{algo:nestedcollectionforspecificROOTclient} on this problem with root node as $\mc{S}_1=\{S_1 = \emptyset\}$ to obtain one possible nested collection.

\begin{itemize}
    \item The following initializations are made at line-2 of the algorithm. $\mc{S}_1=\{S_1\}$, $\lambda=1$, $i=1$.
    \item The \textbf{do-while} loop iterates as long as we collect a nested collection of maximum depth.
    \item \textbf{Iteration-1} of the \textbf{do-while} loop:
    \begin{itemize}[leftmargin=0.05in]
        \item At line-4, $\mc{S}_{temp}$ and $\mc{S}_{2}$ are initialized to $\emptyset$, where $\mc{S}_{temp}$ is the temporary set used to collect the clients at level-2 of nested collection. Once the collection is complete, it is copied into $\mc{S}_{2}$.
        \item The only client in $\mc{S}_1$ is $S_1$, so the \textbf{for} loop at line-5 iterates only once. At line-6 for every $j\notin S_1$ i.e., $j=\{1,2,3,4\}$ we repeat the procedure from lines-7 to 15.
        \begin{itemize}[leftmargin=0.05in]
        \item For $j=1$, $\mc{S}_{1,S_1}$ is collection of all clients  $S^{\prime\prime}\in \mc{S}$ such that $S_1\cup \{1\}=\{1\} \subseteq S^{\prime\prime}$. Thus, $\mc{S}_{1,S_1}=\{S_2,S_4,S_5,S_6,S_7,S_{11}\}$. Now, since $\mc{S}_{1,S_1}\neq \emptyset$, at line-9 we assign a smallest sized client to $S^\prime$. Thus, $S^\prime=S_2$. At line-10 $\mc{S}_{temp}$ is updated as $\mc{S}_{temp}=\emptyset \cup \{S_2\}=\{S_2\}$.
        \item For $j=2$, $\mc{S}_{2,S_1}=\{S_2,S_5,S_6,S_8,S_9,S_{10}\}$ and since $\mc{S}_{2,S_1}\neq \emptyset$, at line-9 we assigns $S^\prime=S_8$ which is the smallest sized among $\mc{S}_{2,S_1}$. At line-10 we update $\mc{S}_{temp}=\{S_2\} \cup \{S_8\}=\{S_2, S_8\}$. Similarly for $j=3$, and $j=4$, $\mc{S}_{3,S_1}=\{S_3,S_5,S_7,S_{10},S_{11}\}$ and $\mc{S}_{4,S_1}=\{S_4,S_6,S_7,S_9\}$. $S^\prime=S_3$ and $S^\prime=S_4$ respectively for $j=3$ and $j=4$. Thus, after this, we have $\mc{S}_{temp}=\{S_2, S_8, S_3,S_4\}$.
        \end{itemize}
    \item Since $\mathsf{MaxDepthReached}=0$, at line-21, $\mc{S}_2$ gets updated as $\mc{S}_2=\mc{S}_{temp}=\{S_2, S_8, S_3,S_4\}$ which are the clients at level-2 in the nested collection. At line-22, $i$ is incremented to $2$. 
    \end{itemize}
\item \textbf{Iteration-2} of the \textbf{do-while} loop:
    \begin{itemize}[leftmargin=0.05in]
        \item Line-4 initializes $\mc{S}_{temp}=\mc{S}_3=\emptyset$.
        \item There are four clients in $\mc{S}_2$, so the \textbf{for} loop at line-5 iterates four times. At the end of these, successors for each client at level-$2$ are found.      

A valid resulting  $\mc{S}_{temp}$ is $$\mc{S}_{temp}=\{S_5,S_6,S_2,S_9,S_{11},S_{10},S_7\}.$$
    \item Since $\mathsf{MaxDepthReached}=0$, at line-21, $\mc{S}_3$ is updated to $\{S_5,S_6,S_2,S_9,S_{11},S_{10},S_7\}$, which are the clients at level-$2$ in the nested collection. At line-22, $i$ increments to $3$. 
    \end{itemize}
\item \textbf{Iteration-3} of the \textbf{do-while} loop:
    \begin{itemize}[leftmargin=0.05in]
        \item Line-4, initializes $\mc{S}_{temp}=\mc{S}_4=\emptyset$.
        \item There are seven clients in $\mc{S}_3$, so the \textbf{for} loop at line-5 has to iterate for seven times. But, $S_5\in\mc{S}_3$, $j=4$ and $S_5 \cup \{4\}=\{1,2,3,4\}\notin \mc{S}$, thus $\mc{S}_{4,S_5}=\emptyset$ which makes line-12 of the algorithm assign $\mathsf{MaxDepthReached}=1$ and $\lambda=3$. This makes the algorithm to come out of the \textbf{for} loop at line-13.
    \item Further, $\mathsf{MaxDepthReached}=1$, at line-16 breaks and comes out of the \textbf{do-while} loop.
    \end{itemize}
\item Line-25 returns the nested collection as $\mc{S}_1=\{\emptyset\}$, $\mc{S}_2=\{S_2, S_8, S_3,S_4\}$, $\mc{S}_3=\{S_5,S_6,S_2,S_9,S_{11},S_{10},S_7\}$ as level-1, level-2 and level-3 clients respectively. It also returns the depth of nested collection as $\lambda=3$. \hfill $\blacksquare$
\end{itemize}
\end{example}

We now calculate the complexity of Algorithm \ref{algo:nestedcollectionforspecificROOTclient}. The crucial step is line 6, which computes successor successor side-information sets in $O(m^2n)$ time. The loops themselves execute at most $mn$ times. Thus, overall the algorithm executes in $O(m^3n^2)$ time. 
 
Note that Algorithm \ref{algo:nestedcollectionforspecificROOTclient} does not guarantee finding an optimal nested collection (i.e., one of maximum length) rooted at the given client, for the PICOD problem given, due to the arbitrary choice of a successor set made in line 8. Executing Algorithm \ref{algo:nestedcollectionforspecificROOTclient} with each client as the root thus enables construction of $n$ nested collections, the largest length of which serves as a lower bound for the nesting number. Clearly, this can be performed in $O(m^3n^2)$ time.
\subsection{Using Theorem \ref{thm:lowerboundeta}: Existing and Novel Results}
\label{subsec:applicationsofnestingnumber}
In this subsection, we use our nested-collections-based lower bound to obtain results for PICOD scenarios with special structural constraints. Firstly, we show a class of problems for which the lower bound in Theorem \ref{thm:lowerboundeta} is tight. Further, we obtain a novel converse bound for the class of complete-$\Sigma$ PICOD problems, for any $\Sigma\subseteq [0:m-1]$ representing the sizes of the side-information sets in the problem. This recovers and further generalizes results from \cite{liu2019tight}.

As part of this subsection, we utilize  Theorem \ref{thm:lowerboundeta} to recover results from \cite{ong2019optimal} for classes of PICOD problems with structural constraints. Though these results are not novel, coupled with Algorithm \ref{algo:nestedcollectionforspecificROOTclient}, they indicate that nested collections can be used for computing first-order lower bounds for PICOD problems. 

Also, the absent-clients-based bounds from \cite{ong2019optimal,ong2019improved} enable characterizations of PICOD problems with few absent clients (thus, for many `present' clients). Specifically, the complete characterization for PICOD solutions for upto four absent clients is given in \cite{ong2019improved}. Theorem \ref{thm:lowerboundeta} can enable the characterization of problems with few present clients. As an illustration, we obtain the complete characterization of PICOD problems with upto three present clients, using the lower bound Theorem \ref{thm:lowerboundeta} as part of the arguments. In the process, we also show a sub-class of problems for which the bounds $\eta(\ch)$ and $\tau_1(\ch)$ can be equally loose, while $\tau_2(\ch)$ is tight. 
\subsubsection{ A class of problems for which $\eta(\ch)$ is a tight lower bound for $\beta(\ch)$}
\label{subsubsec:etaistight}
The following corollary is obvious from \cite[Theorem 1]{subramanian2022bounding} and Theorem \ref{thm:lowerboundeta}. 
\begin{corollary}
\label{Corollary-1}
Let $\mc{H}$ be a PICOD hypergraph with $\eta(\mc{H}) = \D(\mc{H})$. Then, $\beta_q(\mc{H}) = \D(\mc{H})$ for every prime power $q$, and hence $\beta(\ch)=\D(\ch)$.
\end{corollary}
As an application of Corollary \ref{Corollary-1}, we have the following result. The proof of Lemma \ref{lemma:specificclassEtaTight} is available in the Appendix \ref{subsec:ProofspecificclassEtaTight}.
\begin{lemma}
\label{lemma:specificclassEtaTight}
Let $A\subseteq[1:m]$. Consider the PICOD problem $\ch$ with its request sets such that they can be arranged as an ordered list of $L$ subsets denoted by ${\cal R}_i\subset {\cal R}: i\in[1:L]$, with the below structure. 
\begin{enumerate}[leftmargin=0.1in]
    \item The collection ${\cal R}_i$ forms a partition of $A$, for each $i\in[1:L]$.
    \item For each $i\in[1:L-1]$, for each $R\in {\cal R}_i$, there exists a collection of request sets denoted by ${\cal R}_{i+1,R}$, such that ${\cal R}_{i+1,R}\subseteq {\cal R}_{i+1}$ and the subsets in  ${\cal R}_{i+1,R}$ form a non-trivial partition of $R$.
\end{enumerate}    
Then $\beta(\ch)=\D(\ch)=\eta(\ch)=L$. 
\end{lemma}

\subsubsection{A New Proof for a Lower Bound from \cite{liu2017ITW} for Complete-$\Sigma$ PICOD} 
\label{subsubsec:completeSpicod}
The work \cite{liu2019tight} obtained tight converse bounds for a number of classes of PICOD problems with special structure on the side-information sets. These problems fall under the broader class called \textit{complete}-$\Sigma$ PICOD problems. A complete~-$\Sigma$ PICOD problem, defined by the set $\Sigma\subseteq[0:m-1]$, is the PICOD problem consisting of the clients with $\sum_{s\in \Sigma}\binom{m}{s}$ side-information sets, given by ${\cal S}=\{S\subseteq[1:m]: |S|\in \Sigma\}$. For a variety of choices of $\Sigma$, such as a set of consecutive integers or its complement, the work \cite{liu2019tight} characterizes the optimal PICOD($t$) lengths, for every $t$. For PICOD problems with general choices of $\Sigma$, including those containing non-consecutive integers, it was shown in \cite{liu2017ITW} that the optimal PICOD length is at least $|\Sigma|$.  We now provide a new (arguably, simpler) proof, using Theorem \ref{thm:lowerboundeta}, for this existing lower bound known from \cite{liu2017ITW}.
\begin{lemma} \cite[Proposition 1]{liu2017ITW}
    \label{lemma:lowerboundCompleteSproblems}
    For any $\Sigma\subseteq [0:m-1]$, the optimal PICOD length $\beta$ of the complete-$\Sigma$ PICOD problem is lower bounded as $\beta\geq |\Sigma|$.
\end{lemma}
\begin{remark}
    Lemma \ref{lemma:lowerboundCompleteSproblems} subsumes the converse \cite[Theorem 1]{liu2019tight} for complement-consecutive complete-$\Sigma$ PICOD($1$) problems, in which $\Sigma=[0:m-1]\setminus[s_{\mathsf{min}}:s_{\mathsf{max}}]$ for some $0< s_{\mathsf{min}}\leq s_{\mathsf{max}}< m-1$.  However, it is a loose bound for consecutive complete-$\Sigma$ PICOD($1$) problems, in which $\Sigma=[s_{\mathsf{min}}:s_{\mathsf{max}}]$ for $0\leq s_{\mathsf{min}}\leq s_{\mathsf{max}}\leq m-1$. For such problems, the tight bound in \cite[Theorem 2]{liu2019tight} is $\min(s_{\mathsf{max}}+1,m-s_{\mathsf{min}})$, whereas Lemma \ref{lemma:lowerboundCompleteSproblems} suggests $s_{\mathsf{max}}-s_{\mathsf{min}}+1$. These two bounds coincide only when $s_{\mathsf{min}}=0$ or when $s_{\mathsf{max}}=m-1$, otherwise Lemma \ref{lemma:lowerboundCompleteSproblems} gives a strictly loose bound.
\end{remark}

\subsubsection{Recovering Absent-Clients-based Converse Results from \cite{ong2019optimal}}
We now obtain new proofs for all results from \cite{ong2019optimal} for PICOD problems with special structure, using Theorem \ref{thm:lowerboundeta}. 

Note that, in any PICOD problem, the set $[1:m]$ is never present (as no client exists with all messages as side-information), nor is considered to be an absent client (by definition). A sequence $S_1\subsetneq S_2 \subsetneq \hdots \subsetneq S_L$ of $L$ absent clients is said to be an $L$-length nested chain of absent clients. We give an alternative proof for the following lemma, the proof of which is available in the Appendix \ref{subsec:maxlengthnestedabsentProof}.
\begin{lemma} \cite[Lemma 7]{ong2019optimal}
\label{lemma:maxlengthnestedabsent}
For any given PICOD problem $\ch$ on $m$ messages, let $L$ be the maximum length of any nested chain of absent clients in $\ch$. Then $\beta(\ch)\geq m-L$. 
\end{lemma}

We remark that Lemma \ref{lemma:maxlengthnestedabsent} is the basis for another earlier result \cite[Theorem 4]{ong2019optimal}, and thus we obtain the same. We now focus on other results from \cite{ong2019optimal}. 
\begin{lemma}\cite[Theorem 1, converse part]{ong2019optimal}
\label{lemma:unionofabsentRxsNOT0-m-1}
Let $\ch$ be a PICOD problem with side-information sets $\cal S$ such that $\bigcup\limits_{\substack{S\subsetneq [1:m]:\\ S\notin {\cal S}}}S\neq [1:m].$ Then, $\beta(\ch)\geq m-1$. 
\end{lemma}

We complete this part of the present subsection with the following proof for another result from \cite{ong2019optimal}. 
\begin{lemma} \cite[Theorem 3]{ong2019optimal}
\label{lemma:1or0absentnestedpair}
Suppose that the given PICOD problem has a collection of absent clients denoted by ${\cal S}_{abs}$, satisfying one of the following conditions: (a) no pair of clients in ${\cal S}_{abs}$ forms a nested chain of absent clients, or (b) among ${\cal S}_{abs}$, exactly one pair of absent clients forms a nested chain. Then, $\beta(\ch)\geq m-1$. 
\end{lemma}
The proofs of Lemma \ref{lemma:unionofabsentRxsNOT0-m-1} and Lemma \ref{lemma:1or0absentnestedpair} are available in Appendix \ref{subsec:unionofabsentRxsNOT0-m-1Proof} and Appendix \ref{subsec:1or0absentnestedpairProof} respectively.
\subsubsection{PICOD problems containing upto $3$ clients}
\label{subsubsec:uptothreeclients}
We now turn our attention to the characterization of the optimal PICOD lengths for problems with upto three clients. Note that, in \cite{subramanian2022bounding} it was argued that optimal solutions of all problems with degree $\Delta\leq 3$ are characterizable, based on the achievable scheme we presented in \cite{subramanian2022bounding} and prior results from index coding literature. This includes the scenario of upto three clients as well. However, the arguments there depended on enlisting finding suitable demand choices, which satisfy some criteria, across all clients. This may incur high computational complexity. Here, we focus on the PICOD problems upto three clients and characterize their optimal PICOD lengths as well as the associated lower bounds completely, based only on the relationships between the side-information sets. We first consider the case of upto two clients (Lemma \ref{lemma:twoclientcase}, proof in Appendix \ref{subsec:twoclientcaseProof} ) and $n=3$ case (Lemma \ref{lemma:threeclients} , proof in Appendix \ref{subsec:threeclientsproof}). 

\begin{lemma}
\label{lemma:twoclientcase}
    $\beta(\ch)=\tau_2(\ch)=\tau_1(\ch)=\eta(\ch)=1$, if the number of clients $n\leq 2$. 
\end{lemma}

\begin{lemma}
\label{lemma:threeclients}
Let $S_i:i=1,2,3$ be the side-information sets of a PICOD problem $\ch$ containing three clients. Then, the following statements are true, and provide an exhaustive characterization of the parameters $\beta(\ch),\tau_2(\ch),\tau_1(\ch)$ and $\eta(\ch)$. 
\begin{enumerate} [leftmargin=0.1in]
    \item If $\cup_{i=1}^3S_i\neq [1:m]$, then $\beta(\ch)=\tau_2(\ch)=\tau_1(\ch)=\eta(\ch)=1$. 
    \item If $\cup_{i=1}^3S_i=[1:m]$, then $\beta(\ch)=\tau_2(\ch)=\tau_1(\ch)=\eta(\ch)=1$ under either of the following scenarios.
    \begin{enumerate}
        \item There exist distinct $i,j,k$, such that the sets $S_i\setminus(S_j\cup S_k)$ and $(S_j\cap S_k)\setminus S_i$ are non-empty. 
        \item For each choice of distinct $i,j,k$, the set $(S_i\cap S_j)\setminus S_k$ is non-empty.
    \end{enumerate}
    \item There exists a nested collection of length $2$ in $\ch$, only if $S_i\subsetneq S_j$ and $S_i\subsetneq S_k$, for some distinct $i,j,k$.
    \item If there exists a nested collection of length $2$, then $\beta(\ch)=\tau_2(\ch)=\tau_1(\ch)=\eta(\ch)=2$.  
    \item If the following conditions hold: $(i)$ $\cup_{i=1}^3S_i= [1:m]$, $(ii)$ there exists no nested collection of length $2$, and $(iii)$ neither of the conditions in \textit{(2a)} and \textit{(2b)} hold, then $\beta(\ch)=\tau_2(\ch)=2$, while $\tau_1(\ch)=\eta(\ch)=1$.
\end{enumerate}
\end{lemma}
\section{Conclusion}
\label{sec:discussion}

In this work, we have presented new converse results for PICOD based on its structure.  Our converse bound based on the nested collections enables us to prove some novel results for PICOD problems with special structure. Further, inspite of this bound not being stronger than those in earlier work, we show advantages in its computability and also obtain new proofs for a number of existing results using this new bound. Future work would include tuning the present method to obtained tighter lower bounds than those in \cite{ong2019improved,ong2019improved}.  Extending this method to PICOD($t$) problems can be an interesting future direction. 




\bibliographystyle{IEEEtran}
\bibliography{IEEEabrv, root.bib}


\begin{appendices}
\section{Proof of Theorem \ref{thm:lowerboundeta}}

\label{subsec:proofoftheoremlowerbound}
In this subsection, we prove Theorem \ref{thm:lowerboundeta} by utilizing the existing converse from \cite{ong2019optimal}. Since it is already known from the results in \cite{ong2019improved} that $\beta(\ch)\geq \tau_2(\ch)\geq \tau_1(\ch)$, it is enough to show that $\tau_1(\ch)\geq \eta(\ch)$. 

Consider a nested collection of length $L$ of $\ch$, denoted by ${\cal S}_i:i\in[1:L]$, where ${\cal S}_i$ denotes the set of side-information sets at level-$i$ of the nested collection. 
To complete the proof of Theorem \ref{thm:lowerboundeta}, we first prove that $\tau_1(\ch)\geq L$. Invoking the definition of the nesting number, the proof will then be complete. To show $\tau_1(\ch)\geq L$, we first trim the given collection to find one that suits our purpose, in a sequential manner, as follows. 
\begin{itemize}
    \item Consider $S_1\in{\cal S}_1$.  Trim ${\cal S}_1$ as ${\cal S}_1=\{S_1\}$. 
    \item For each $i\in[1:L-1]$, 
    \begin{enumerate}
        \item For each $S_i\in{\cal S}_i$ and $j\notin S_i$, let $S_{i+1}(S_i,j)$ denote a side-information set in ${\cal S}_{i+1}$ such that the following are true: (a) $ S_i\cup\{j\}\subseteq S_{i+1}(S_i,j)$, and (b) there exists no $H'\in{\cal S}_{i+1}$ such that both (a) is true and $|H'|> |S_{i+1}(S_i,j)|$.
        \item Trim ${\cal S}_{i+1}$ as follows.
        $${\cal S}_{i+1}=\{S_{i+1}(S_i,j):\forall j\notin S_i,\forall S_i\in{\cal S}_i\}.$$
    \end{enumerate}
\end{itemize}
Essentially, the trimmed collection includes only the largest successors  at level-$(i+1)$ of the hyperedges at level-$i$. Further, the new trimmed collection continues to be a nested collection of length $L$, as the requirements of Definition \ref{defn:nestingcollection} continue to hold as is. Let $\ch'$ denote a new PICOD problem consisting only the clients whose side-information sets are in the trimmed nested collection. Since any solution for $\ch$ is a solution for $\ch'$ as well, we must have $\tau_1(\ch)\geq \tau_1(\ch')$. Thus, showing $\tau_1(\ch')\geq L$ completes the proof of Theorem \ref{thm:lowerboundeta} as well.

We now show $\tau_1(\ch')\geq L$, for which we shall use its definition in \eqref{eqn:oldconversebounds}. Let $D'$ be some decoding choice function for $\ch'$. We construct a ordered set $C_{D'}$ based on an ordering of the messages $[1:m]$, and show that the number of clients hit by $C_{D'}$ (denoted by $\tau(C_{D'})$) is at least $L$.

Recall that $S_1$ is the only client at level-$1$. Let $S'_1=S_1$. Recursively, for $i\in[1:L'-1]$, let $S'_{i+1}$ be the smallest client in $\ch'$, such that $S'_{i+1}$ contains $S'_i\cup\{D'(S'_i)\}$. Here $L'$ is the largest value of $(i+1)$ such that such a pair of clients $(S'_i,S'_{i+1})$ exists. Note that all the $L'$ clients in this sequence are distinct, by definition. Further, we also observe that, by the property that all clients of $\ch'$ are part of the nested collection, each $S'_i$ has a successor in its next higher level. Such a successor serves as a candidate for $S'_{i+1}$. By this observation, we see that $L'\geq L$. However, it is not necessary for $S'_{i+1}$ to be in level-$(i+1)$; instead, $S'_{i+1}$ could be present at a level prior to level-$(i+1)$. 

Now, let $C_{D'}$ be any ordering of message indices $[1:m]$, with the following condition.
\begin{itemize}
    \item For each $i\in[1:L']$, we have $C_{D',|S'_i|}=S'_i$ (i.e., the first $|S'_i|$ elements of $C_{D'}$ form $S'_i$), and $C_{D',|S'_{i+1}|}\setminus C_{D',|S'_i|}=D'(S'_i)$. 
\end{itemize}
We note that there is at least one such ordering, by definition of $S'_i:i\in[1:L']$. 
Further, we also note that this is a valid sequence that follows \textit{Rule A} and \textit{Rule B}. This is because, (a) \textit{Rule A} is followed in the transition from $C_{D',|S'_i|}$ to $C_{D',|S'_{i+1}|}$ for each $i\in[1:L']$, and (b) \textit{Rule B} is followed at every other point in the sequence.  
Here (b) is true, since (i) for each $i\in[1:L'-1]$, there exists no client in $\ch'$ with a side-information set of size strictly smaller than $|S'_{i+1}|$ that also includes $S'_i\cup D(S'_i)$, by construction of the trimmed nested collection, and because (ii) there is no client in $\ch'$ which includes $S'_{L'}\cup D(S'_{L'})$ as a proper subset, by definition of $L'$. 

Thus, we observe that there are $L'\geq L$ clients hit by this ordered set $C_{D'}$. Thus, $\tau(C_{D'})\geq L'\geq L$. By invoking the definition of $\tau_1(\ch')$ given in \eqref{eqn:oldconversebounds}, and since $D'$ is chosen arbitrarily, we have that $\tau_1(\ch')\geq L$. This completes the proof of Theorem \ref{thm:lowerboundeta}, as $\beta(\ch)\geq \tau_2(\ch)\geq \tau_1(\ch)\geq\tau_1(\ch')$ and since we can choose our initial nested collection to be the one with the largest length $\eta(\ch)$. 

\section{Proof of Lemma \ref{lemma:specificclassEtaTight}}
\label{subsec:ProofspecificclassEtaTight}

Firstly, we observe that $\Delta(\ch)=L$. This is true, because each message appears in exactly $L$ unique request sets, one in each of the collections ${\cal R}_i:i\in[1:L]$, by the given structure.

Further, the given structure is in fact represents a $L$-length nested collection. This is because, for each $R\in{\cal R}_i$, for each $j\in R$, we have some request set $R'$ in ${\cal R}_{i+1}$ which is a subset of $R\setminus\{j\}$, as per given condition  \textit{2)}. When expressed in terms of side-information sets, this is the criterion required in Definition \ref{defn:nestingcollection}, in terms of the request sets. Thus, $\eta(\ch)= L$.  
Using Corollary \ref{Corollary-1}, we have the claim.
\section{Proof of Lemma \ref{lemma:lowerboundCompleteSproblems}}
\label{subsec:lowerboundCompleteSproblemsProof}

Let $\Sigma=\{s_1,\hdots,s_{|\Sigma|}\}$, where $s_i\leq s_{i+1}$, for all $i\in[1:|\Sigma|-1]$. Consider the ordered list of subsets ${\cal S}_i=\{S\subset[1:m] :|S|\in s_i\}$, for $i\in[1:|\Sigma|]$. It is easy to see that this is a nested collection in the given problem, whose length is $|\Sigma|$. The lemma thus follows by applying Theorem \ref{thm:lowerboundeta}. 
\section{Proof of Lemma \ref{lemma:maxlengthnestedabsent}}

\label{subsec:maxlengthnestedabsentProof}
Let $D$ be some demand function for all clients in $\ch$. Consider any $m$-length sequence of subsets of $[1:m]$ given as $S_0\subsetneq S_1\subsetneq \hdots\subsetneq S_{m-2}\subsetneq S_{m-1}$, where $|S_i|=i, \forall i\in[0:m-1]$, such that $S_{i+1}\setminus S_i=D(S_i)\in[1:m]$ (i.e., the demand of client $S_i$), if $S_i$ is present in $\ch$. Due to the given statement, there are at least $m-L$ present clients within this sequence. Let these, in sequence, be $S_{l_1}\subsetneq S_{l_2}\subsetneq \hdots \subsetneq S_{l_{m-L}}$. 

Consider the execution of Algorithm \ref{algo:nestedcollectionforspecificROOTclient}, with the root client being $S_{l_1}$. Observe that, in this execution, we have by the above observations, the following sets (as defined in line 7 of Algorithm \ref{algo:nestedcollectionforspecificROOTclient}) are non-empty: 
\begin{align*}
    {\cal S}_{D(S_{l_k}),S_{l_k}}:~ k\in [1:m-L-1]. 
\end{align*}
Since this is true for each choice of the demand function $D$, we thus see that the nested collection, rooted in $S_{l_1}$, obtained according to Algorithm \ref{thm:lowerboundeta}, has length $m-L$. Applying Theorem \ref{thm:lowerboundeta} then completes the proof. 
\section{Proof of Lemma \ref{lemma:unionofabsentRxsNOT0-m-1}}
\label{subsec:unionofabsentRxsNOT0-m-1Proof}
By the given condition, we see that there exists at least one index $a\in[1:m]$ such that the set $\{a\}$ is a present client. Now, consider the collection defined as follows: for each $i\in[1:m-1]$, let ${\cal S}_i$ be the set of all $i$-sized subsets of $[1:m]$ which contain index $a$. All these clients are clearly present, by the given condition. Further, it is easy to see that ${\cal S}_i:i\in[1:m-1]$ is a nested collection, which has length $m-1$. Invoking Theorem \ref{thm:lowerboundeta} completes the proof.

\section{Proof of Lemma \ref{lemma:1or0absentnestedpair}}
\label{subsec:1or0absentnestedpairProof}

If the condition (a) holds, then the claim follows directly from Lemma \ref{lemma:maxlengthnestedabsent}, as the length of the longest nested chain of absent clients is at most $1$, in this case.

Now consider the case when (b) holds. Let $A$ and $B$ be the absent clients such that $A\subsetneq B$. Suppose $\emptyset$ is an absent client. Then, by the given condition, it is clear that $A=\emptyset$. Note that $[1:m]\setminus B$ is guaranteed to be non-empty. For any $a\in[1:m]\setminus B$, consider the collection ${\cal S}_i:i\in[1:m-1]$ where ${\cal S}_i$ consists of all $i$-sized subsets of $[1:m]$ which contain $a$. Note that all clients in this collection are present and further, this is a nested collection of length $m-1$. Invoking Theorem \ref{thm:lowerboundeta} completes the proof in this case.

Now, consider the case when (b) holds and $\emptyset$ is a present client. We will assume that  $A=\{m-|A|+1,\hdots,m\}$ and $B=\{m-|B|+1,\hdots,m\}$, without loss of generality (since we can always achieve this by applying a suitable permutation on the indices $[1:m]$). Note that $1\notin B$, as $|B|\leq m-1$ (true for any side-information set). Consider a  problem $\ch'$ modified from $\ch$, with the only difference from $\ch$ being that the set $B$ is a present client in $\ch'$. Thus, in $\ch'$, the length of the longest nested chain of absent clients is $1$. Consider the nested collection of $\ch'$, generated by a slightly modified version of Algorithm \ref{algo:nestedcollectionforspecificROOTclient}, rooted at the client $\emptyset$. The modification we introduce is as follows. Instead of executing line $9$ of Algorithm \ref{algo:nestedcollectionforspecificROOTclient} as is (to pick the successor of client $S$), we execute
\begin{align}
\label{eqn:newwaytoaddtocollection} S'\gets \arg\min\limits_{S''\in \mc{S}_{j,S}} \left(\sum_{s\in S''}s\right).
\end{align}
That is, we choose $S'$ as that present client whose element-wise sum is minimum among all those that contain $S\cup\{j\}$. Note that this condition makes the choice of $S'$ unique. Also, the arguments in the proof of Lemma \ref{lemma:maxlengthnestedabsent} continue to hold for this modified algorithm as well. Hence, by the proof of Lemma \ref{lemma:maxlengthnestedabsent}, this nested collection of $\ch'$ has $m-1$ levels. We seek to show that, removing $B$ from all its occurrences in this collection, does not change its property as a nested collection with $m-1$ levels. To do this, we first show that $|B|$ is present only at level-$(|B|+1)$ of the nested collection (only if $|B|+1<m$), otherwise it is not present at all.

Firstly, note that, for all orderings of the elements of $B$ such that the subset with the first $|A|$-elements is not equal to $A$, each subset of $B$ comprising of the first $k$ elements in that order (corresponding to $k\in[1:|B|-1]$) is present, since no two distinct nested pairs of clients are absent. Thus, every proper subset $B'\subseteq B$ such that $|B'|\neq A$ is present in level-$(|B'|+1)$ of the nested collection. In particular, the set $B\setminus\{b\}$, for any $b\in B$ such that $B\setminus\{b\}\neq A$, is present in the nested collection at level-$|B|$. As the successor of any such set, $B$ is included in the nested collection at level-$(|B|+1)$, only if $|B|< m-1$. Note that if $|B|=(m-1)$, then this cannot happen, as the nested collection itself terminates at level-$(m-1)$. 

Now, if at all $B$ occurs in the nested collection, we show that $B$ does not occur in  in any other level apart from level-$(|B|+1)$. By the construction of any nested collection rooted at $\emptyset$, note that a set of size $|B|$ can only occur in the collection at level-$i$, for some $i\leq |B|+1$. Now, if $B$ has to occur at a level-$i$ where $i<|B|+1$, then there has to be a proper subset of $B$ (say $B_1$) for which a successor $B_2$ (as per \eqref{eqn:newwaytoaddtocollection}) exists, such that both of the following conditions hold (i) $B_2\subseteq B$, and (ii) $|B_2|-|B_1|>1$. We now show that no such $B_1,B_2$ exists.

For any $B_1$ such that either (a) $|B_1|\in[0:|A|-2]\cup[|A|:|B|-2]$, or (b) $|B_1|=|A|-1$ and $B_1\setminus A\neq \emptyset$ with $|B|=|A|-1$, we have already verified that the successor $B_2$ as per \eqref{eqn:newwaytoaddtocollection} satisfies $|B_2|-|B_1|=1$. Thus, we need to check only the case when $B_1=A\setminus \{a\}$ for some $a\in A$. Note that the set $A\setminus\{a\}$, is present and included in the level-$|A|$ of the nested collection by the given condition. 

Consider the set $A\cup\{1\}$. This is a super-set of $A$. Hence, by the given condition that no two distinct nested pairs of clients are absent, the set $A\cup\{1\}$ must be present. Thus, for each $a\in A$, the set $A$ being absent, the modified execution \eqref{eqn:newwaytoaddtocollection} implies that the chosen successor of $A\setminus\{a\}$ with respect to $a$ at level-($|A|+1$)  will be the client $A\cup\{1\}$, thus satisfying condition (ii). However, note that $A\cup\{1\}$ is not a subset of $B$, thus it does not satisfy condition (i). Hence, there is no $B_1, B_2$ that satisfy both conditions required. Hence $B$ occurs only at the level-$(|B|+1)$ in the nested collection, and further only if $|B|<m-1$. 

Finally, we argue that removing $B$ does not affect the nested-collection property. To see this, we should show that any set $B\setminus b$ (apart from $A$, if $|A|=|B|-1$) in level-$|B|$ has another successor in level-$(|B|+1)$, apart from $B$. We now argue this is the case. 

By prior arguments and because of the given condition, all sets containing  $A\cup\{1\}$ must also be present. 
In particular, the set $B\cup\{1\}$ will be present at level-$(|B|+1)$. This serves as a successor to  $B\setminus\{b\}$ at level-$|B|$. 
Thus, the set $B$ can be removed from the collection, while retaining its identity as a nested collection. Note that this new nested collection without $B$ is a valid nested collection in our original problem $\ch$. This completes the proof. 

\section{Proof of Lemma \ref{lemma:twoclientcase}}
\label{subsec:twoclientcaseProof}
    Trivially, one transmission is always sufficient when the number of clients $n=1$. We now consider the $n=2$ case. It is enough to show $\beta(\ch)=1$, as for any problem, it is easy to see that each converse bound is at least $1$. 
    
    Suppose the side-information sets are such that the sets $S_1\setminus S_2$ and $S_2\setminus S_1$ are non-empty. Then, the transmitter can choose a message $b\in S_1\setminus S_2$ and another $b'\in S_2\setminus S_1$, and transmits the sum $b+b'$. This satisfies both the clients. Otherwise, if $S_i\subsetneq S_j$ for some $i,j$, then since $|S_j|\leq m-1$, the transmitter can transmit some $b\in[1:m]\setminus S_j$, which satisfies both clients. 


\section{Proof of Lemma \ref{lemma:threeclients}}
\label{subsec:threeclientsproof}

Observe that, from \cite[Proposition 8]{liu2019tight}, $\beta(\ch)=1$ if and only if the demand choices for the three clients form an independent set of $\ch$, i.e., no two of the demand choices can be simultaneously present in the request set of any one of the three clients. Consider such demand choices for each client, denoted by $b_{d_1},b_{d_2},b_{d_3}$, where $d_i\in[1:m]$. If $d_1=d_2=d_3$, then this means $d_1\notin \cup_{i}S_i$, as otherwise at least one client will not be satisfied by this choice. This gives part \textit{(1)} of the claim. If $d_1= d_2\neq d_3$, then it should be the case that $d_1\notin S_1\cup S_2$ and $d_3\notin S_3$. By Lemma \ref{lemma:lengthone}, $d_3\in S_1\cap S_2$, and also $d_1\in S_3$. This gives part \textit{(2a)} of the claim. Finally, if $d_1,d_2,d_3$ are all distinct, then $d_i$ must be in $(S_j\cap S_k)\setminus S_i$, for all distinct $i,j,k$, again by Lemma \ref{lemma:lengthone}. This concludes the proof of part $(2b)$. Also, we note that in all cases when $\beta(\ch)=1$, the lower bounds $\tau_2(\ch),\tau_1(\ch)$ and $\eta(\ch)$ too must be exactly $1$ (as each of them is at least $1$ for any problem). 

Now, suppose that a nested collection of length $2$ exists in $\ch$. Consider a subset $S_i$ in level-$1$. Then, it must have successor-clients (super-sets of $S_i$) in level-$2$, which are of size at most $m-1$. Thus, $S_i$ itself must be of size at most $m-2$, and therefore must have at least two successors in level-$2$ (corresponding all messages not in $S_i$). Thus, it must be the case that in any nested collection of length $2$, level-$1$ has one client and level-$2$ has two. This proves \textit{(3)}. Note that this also shows that the nesting number is $2$. 

For a nested collection of length $2$, let the level-$1$ client be $S_i$ and the level-$2$ be $S_j,S_k$ for distinct $i,j,k$. The clients $S_j,S_k$ can both be satisfied with one transmission as per Lemma \ref{lemma:twoclientcase}. To satisfy $S_i$ one additional transmission (any message not in $S_i$) is needed. This scheme, along with \eqref{eqn:oldconversebounds}, completes the proof of \textit{(4)}. 

Finally, we prove \textit{(5)}. By the conditions \textit{(i)} and \textit{(iii)}, we see that $\beta(\ch)>1$ in this case. Thus, by arguments similar to \textit{(4)}, $\beta(\ch)=2$. We now show $\tau_2(\ch)=2$. 


Firstly, we observe that under the conditions in case \textit{(5)} there exists no distinct $i,j$ such that $S_i\subset S_j$. Indeed, suppose this was true, then we see that the problem reduces to satisfying only two clients, $S_j$ and $S_k$ ($k\neq i,j$), which means one of the conditions under cases \textit{(1)} and \textit{(2)}  must hold, by prior arguments. However, we have explicitly excluded these conditions in case \textit{(5)}. Thus, $S_i\subset S_j$ is not true for any distinct $i,j$. 

\textit{Case} \textit{(5a)}: In this sub-case, suppose that $(S_i\cap S_j)\setminus S_k$ is empty for all distinct $i,j,k$. 


Now, suppose $S_i\subseteq S_j\cup S_k$ for distinct $i,j,k$.  Then, any $x\in S_i\setminus S_k$ (such an element exists as $S_i\setminus S_k$ is non-empty) also lies in $S_j$. Thus, $x\in S_i\cap S_j\setminus S_k$, giving a contradiction to \textit{(5a)}. Thus, $S_i\setminus(S_j\cup S_k)\neq \emptyset$, for any distinct $i,j,k$ in this case. 

To prove $\tau_2(\ch)=2$, we will show that for every choice of demands $D$, we can get an ordered set $C_D$ of $[1:m]$ according to the rules in \cite{ong2019improved} (summarized in Subsection \ref{subsec:conversesabsentreceiver}) such that the number $\tau'(C_D)$ of non-skipped messages in $C_D$ (following either \textit{Rule A} or \textit{Option B'}) is $2$. 

Now, by condition \textit{(i)}, for any demand choices $D$, for each $i$, we must have $D(S_i)=d_i\in S_{i'}\setminus S_i$, for some $i'\neq i$. Further, by \textit{(5a)}, for each $i$, we must have that $D(S_i)\in S_{j}\setminus (S_i\cup S_k)$, where $j,k,i$ are all distinct. Because of this, it must be the case that there exist distinct $i,j,k$ such that $D(S_i)=d_i\in S_{j}\setminus (S_i\cup S_k)$ \textit{and} $D(S_j)=d_j\in S_{k}\setminus (S_j\cup S_i)$. 

Now, we construct a $C_D$ as follows.
\begin{itemize}
    \item Let $S_i$ be the first client that is hit by $C_D$, i.e., $C_{D,|S_i|}=S_i$. Note that we can always skip enough indices such that this is true. Then, we will have $C_{D,|S_i|+1}=S_i\cup\{d_i\}$. Thus, $d_i\in S_j\setminus (S_i\cup S_k)$ is a non-skipped message index. 
    \item Subsequently, let $S_j$ be the next client that is encountered, i.e., $C_{D,|S_i\cup S_j|}=S_i\cup S_j$. Note that we can always skip indices such that this is true. Then, we will again have to add the non-skipped message index $D(S_j)=d_j\in S_k\setminus (S_i\cup S_j)$, to obtain, $C_{D,|S_i\cup S_j|+1}=S_i\cup S_j\cup\{d_j\}$. 
    \item Finally, we skip other messages not added so far, until we encounter $S_k$, at which point we will still skip $D(S_k)$, as $D(S_k)\in (S_i\cup S_j)$ for sure. 
\end{itemize}
Thus, we observe that, for any demand choice function $D$, we must have some choice for $C_D$ such that $\tau'(C_D)=2$. Now, from \eqref{eqn:oldconversebounds}, we see that $\tau_2(\ch)=2$. 

We now argue that $\tau_1(\ch)=1$. To see this, note that  $S_i\setminus S_j$ is non-empty for all distinct $i,j$. Thus, if we hit any client $S_i$ while constructing $C_D$ as per \textit{Rule A} and \textit{Rule B}, we cannot hit another client $S_j, j\neq i$. Thus, $\tau_1(\ch)=1$. Finally, $\eta(\ch)=1$ as there is a trivial nested collection consisting of any one client. This completes the proof of \textit{5a)}.




Case \textit{(5b)}: In this sub-case, consider that there exists distinct $i,j,k$ such that $(S_j\cap S_k)\setminus S_i$ is non-empty, but for any such case $S_i\setminus (S_j\cup S_k)$ is empty. Without loss of generality, assume that $S_1\subseteq (S_2\cup S_3)$ whenever $(S_2\cap S_3)\setminus S_1$ is non-empty.

In this case, let us fix the demand choices $D$ such that $D(S_1)=d_1\in (S_2\cap S_3)\setminus S_1$, $D(S_2)=d_2\in S_1\setminus S_2$ and $D(S_3)=d_3\in S_1\setminus S_3$ (note that both of these sets must be non-empty, as we have argued). As $S_1\subseteq (S_2\cup S_3)$, we must thus have $d_2\in S_3$ and $d_3\in S_2$. Thus, we see that $S_1\cap S_3\setminus S_2$ and $S_2\cap S_3\setminus S_1$ are non-empty as well. This implies that the condition in case \textit{(2b)} holds, which is a contradiction as we assumed in condition \textit{(iii)} of case \textit{(5)}. 

Note that there are no further cases to be considered under case \textit{(5)}. The fact that this provides an exhaustive characterization of the parameters $\beta(\ch),\tau_2(\ch),\tau_1(\ch)$ and $\eta(\ch)$ follows from the conditions in the statements, which provide an exhaustive categorization of all problems with three clients. This concludes the proof of the claim.

\end{appendices}

\end{document}